# Valley pseudospin in monolayer $MoSi_2N_4$ and $MoSi_2As_4$

Chen Yang,[1,2,†] Zhigang Song,[3,†,*] Xiaotian Sun,[4] Jing Lu[1,2,5,6]

[1]State Key Laboratory for Mesoscopic Physics and Department of Physics, Peking University, Beijing 100871, P. R. China
[2]Academy for Advanced Interdisciplinary Studies, Peking University, Beijing 100871, P. R. China
[3]Computational Research Division, Lawrence Berkeley National Laboratory, Berkeley, CA 94720, USA
[4]College of Chemistry and Chemical Engineering, and Henan Key Laboratory of Function-Oriented Porous Materials, Luoyang Normal University, Luoyang 471934, P. R. China
[5]Collaborative Innovation Center of Quantum Matter, Beijing 100871, P. R. China
[6]Beijing Key Laboratory for Magnetoelectric Materials and Devices, Beijing 100871, P. R. China

*Corresponding author: Zhigang Song (szg@pku.edu.cn)
†These authors contributed equally to this work

**Abstract:** For a long time, two-dimensional (2D) hexagonal $MoS_2$ was proposed as a promising material for the valleytronic system. However, the limited size of growth and low carrier motilities in $MoS_2$ restrict its further application. Very recently, a new kind of hexagonal 2D MXene, $MoSi_2N_4$, was successfully synthesized with large size, excellent ambient stability, and considerable hole mobility. In this paper, based on the first-principles calculations, we predict that the valley-contrast properties can be realized in monolayer $MoSi_2N_4$ and its derivative $MoSi_2As_4$. Beyond the traditional two-level valleys, the valleys in monolayer $MoSi_2As_4$ are multiple-folded, implying a new valley dimension. Such multiple-folded valleys can be described by a three-band low-power Hamiltonian. This study presents the theoretical advance and the potential applications of monolayer $MoSi_2N_4$ and $MoSi_2As_4$ in valleytronic devices, especially multiple information processing.

## Introduction

Valleytronics has aroused great attention in recent years.[1-5] In valleytronics, the wave envelope in real space is locked to the special crystal momentum in reciprocal space, resulting in the tunable electronic and optical properties. Several exciting phenomena, such as the valley Hall effect,[6-8] valley excitons,[9-11] and the valley Zeeman effect,[12-14] have been achieved in the previous experiments. The semiconducting two-dimensional (2D) transition-metal-dichalcogenide, for example, $MoS_2$, has been proposed as an ideal material for the potential valleytronic system for a long time,[15-22] and the previous researches on valleytronics mainly focused on the $MoS_2$ family.[23-28] However, the limited size of growth and low carrier motilities restrict 2D $MoS_2$ for further transports applications. This encourages researchers to discover new promising valleytronic materials, such as 2D honeycomb BN, $VAgP_2Se_6$, $CuCrP_2Se_6$, $HfN_2$, $Nb_3I_8$, and so on.[29-33] 2D valleytronic materials, which inherit and surpass the $MoS_2$ family, are expected to enhance the variety of valleytronic materials. Especially, the stable and easily synthesized candidates are desired.

Very recently, a new kind of hexagonal 2D MXene, 2D $MoSi_2N_4$, has been successfully synthesized with a large size of up to 15 mm × 15 mm.[34] The ambient stability of monolayer $MoSi_2N_4$ is excellent and could present no structural change of samples under ambient conditions for even six months. The elastic modulus of monolayer $MoSi_2N_4$ is four times as big as that in monolayer $MoS_2$.[34-35] Besides, the large theoretical electron/hole motilities are up to ~ 270/1200 $cm^2$ $V^{-1}$ $s^{-1}$ in monolayer $MoSi_2N_4$, which are also four to six times larger than those of monolayer $MoS_2$.[34, 36] The high mobility and stability in the experiment promise an advantage in valley transports. Finally, monolayer $MoSi_2N_4$, and its derivatives, for example, monolayer $MoSi_2As_4$ also have appropriate bandgaps in a wide range from direct 0.60 eV to indirect 1.73 eV, respectively, implying potential optical applications in the visible range.[34] In this paper, based on the first-principles calculations, we predict that hexagonal monolayer $MoSi_2N_4$ and its derivatives have a pair of valley pseudospins similar to monolayer $MoS_2$ and even show the potential to surpass monolayer $MoS_2$ in several valleytronic properties, for example, multiple-folded valleys in monolayer $MoSi_2As_4$. This study presents promising candidates for multiple-folded valleys potential to be applied in multiple information processing in the future.

## Results

The previous study had confirmed the structure of hexagonal 2D $MoSi_2N_4$ through the TEM images combined with the first-principles calculations.[34] As a result, the structure of monolayer $MoSi_2N_4$ could be regarded as a $MoN_2$ layer (similar to $MoS_2$ layer) sandwiched by two Si-N bilayers. In our calculation, as shown in Fig. 1. (a), the optimized lattice constants of monolayer $MoSi_2N_4$ are $a = b = 2.91$ Å, which are consistent with the previous report and a little smaller than those of 3.16 Å in monolayer $MoS_2$.[34, 37-38] Here, we take monolayer $MoSi_2As_4$ as an example of its derivatives. In monolayer $MoSi_2As_4$, arsenic atoms are replaced with nitric atoms. After full optimization, the lattice constants of monolayer $MoSi_2As_4$ are $a = b = 3.62$ Å. All these materials hold $C_3$ rotation symmetry, the inversion asymmetry and a mirror paralleling to the material plane.

The $MoSi_2N_4$ and $MoSi_2As_4$ monolayers are semiconductors with a bandgap of 1.73 and 0.60 eV, respectively, without spin-orbit coupling (SOC) in the previous report.[34] The bandgap of monolayer $MoSi_2N_4$ approximates that of 1.66 eV in monolayer $MoS_2$.[17, 37-38] The phonon spectra of monolayer $MoSi_2N_4$ and $MoSi_2As_4$ are calculated, as shown in Fig. 1. (b) and (c). Herein, the absence of an imaginary mode in the entire Brillouin zone (BZ) confirms the dynamical stability of both hexagonal monolayer $MoSi_2N_4$ and $MoSi_2As_4$. Especially, the phonon spectrum implies $MoSi_2As_4$ can be finally synthesized, although $MoSi_2As_4$ has not been synthesized in experiments. In monolayer $MoSi_2N_4$, short-wave phonon energy is larger than that of monolayer $MoS_2$.[37] This property implies lower thermal scattering in high temperatures of monolayer $MoSi_2N_4$, which is critical to suppress the inter-valley depolarization.

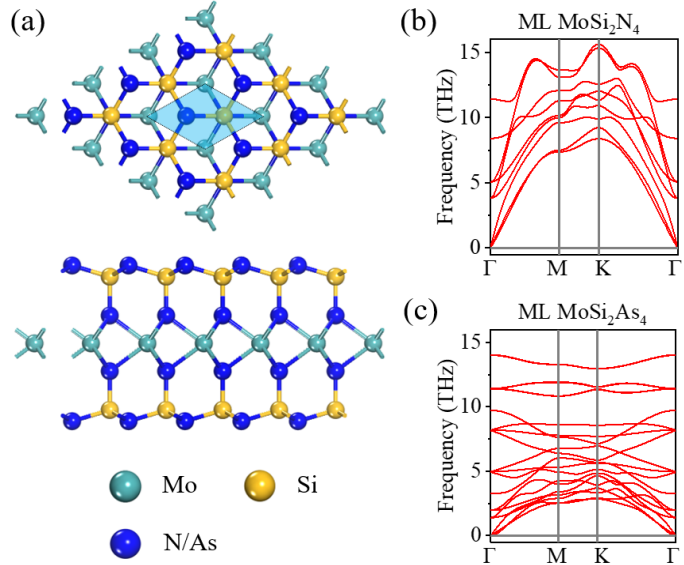

Fig. 1. (a) Top- and side-viewed atomic structures of monolayer $MoSi_2N_4$ and $MoSi_2As_4$. The shadow demonstrates one unit-cell. Phonon spectra for (b) monolayer $MoSi_2N_4$ and (c) $MoSi_2As_4$.

The 3D band structures of monolayer $MoSi_2N_4$ and $MoSi_2As_4$ are shown in Fig. 2. Both hexagonal monolayer $MoSi_2N_4$ and $MoSi_2As_4$ have two valley-like or peak-like bands in the vicinities of the K and K' points, corners of the hexagonal BZ. The next discussion will show that the two valleys are inequivalent due to inversion asymmetry and time-reversal symmetry, although they have degenerate energy. In the middle of Fig. 2. (a) and (b), the momentum-resolved degree of circular polarization figures in both monolayer $MoSi_2N_4$ and $MoSi_2As_4$ are demonstrated, respectively. Mathematically, it has a definition of:

$$\eta(k) = \frac{|P_+(k)|^2 - |P_-(k)|^2}{|P_+(k)|^2 + |P_-(k)|^2} \tag{1}$$

where $P_\pm = P_x \pm P_y$, and $P_\alpha$ is the momentum matrix element between the valence and conduction bands. Due to the protection of the $C_3$ symmetry, the circular dichroism is perfectly valley-selective. A right-handed/left-handed circularly polarized photon can be selectively absorbed around the K/K' point, respectively, with the right-handed/left-handed one wholly prohibited. Then, an electron is excited from the valence band to the conduction band. Therefore, in two inequivalent valleys, circularly polarized optical pumping could selectively generate electrons and holes in only one desired valley, resulting in valley pseudospin polarization.

In monolayer $MoSi_2N_4$, the bandgap is 2.06 eV at K/K' points (without SOC),[34] implying that either valley can be selectively excited by a circularly-polarized phonon with the phonon energy of approximately 2.06 eV. Thus, the electron or hole population of either valley can be controlled. Based on Fig. 2. (b), the optical selection rule is similar in monolayer $MoSi_2As_4$. Besides, monolayer $MoSi_2As_4$ has a direct bandgap, and the conduction band minima (VBM) and valence band maxima (CBM) are well localized at K and K' without states of the same energy from other parts of BZ. This means that the inference from the other of BZ is minimal in monolayer $MoSi_2As_4$. Thus, monolayer $MoSi_2As_4$ has a pair of perfect valleys, even beyond ones in monolayer $MoS_2$. Of course, the advantage of monolayer $MoSi_2As_4$ over $MoS_2$ monolayer is not only this, for example, multi-folded valleys in the following part.

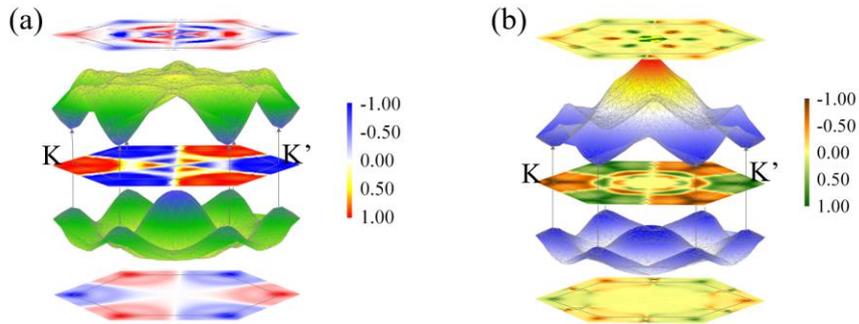

Fig. 2. Valley-contrasting properties of (a) monolayer $MoSi_2N_4$ and (b) $MoSi_2As_4$. Top valence bands and bottom conduction bands are shown in the 3D figures. In the middle, the hexagon is the BZ color-coded by the degree of circular polarization of excitation from the valence band to conduction. The Color bars are for the degree of circular polarization. Berry curvatures of the conduction bands and valence bands are also presented in the top and bottom. The dashed hexagons imply the first BZ.

Berry curvature is like a significant magnetic field in the momentum space. This is the stemming of the different behavior of different valleys in transports. The integration of Berry curvature is the Hall conductance. Mathematically, the Berry curvature can be calculated based on the following expression of:

$$\Omega(k) = i\nabla_k \times \langle u(k)|\nabla_k|k(k)\rangle \quad (2)$$

where $|u(k)\rangle$ is the periodic part of the Bloch wave. The Berry curvature maps of the valence band and the conduction band in monolayer $MoSi_2N_4$ and $MoSi_2As_4$ are shown in the bottom and top of Fig. 2 (a) and (b), respectively. Obviously, Berry curvature of the valence band (and the conduction band) for monolayer $MoSi_2N_4$ and $MoSi_2As_4$ is opposite in two different valleys. A transverse velocity of the valley wave envelope is generated due to this nonzero Berry curvature, if an external in-plane electric field is applied. The transverse velocity induced by Berry curvature would lead to a spatial separation of the carriers coupled to two different valleys. The carriers from different valleys would be accumulated at opposite sides, resulting in a valley Hall effect. Till now, our results demonstrate that the monolayer $MoSi_2N_4$ and $MoSi_2As_4$ offer similar properties to classical valley physics of monolayer $MoS_2$.

To furtherly investigate the valleys in monolayer $MoSi_2N_4$ and $MoSi_2As_4$ and possible properties beyond monolayer $MoS_2$, the fat band structures of monolayer $MoSi_2N_4$ and $MoSi_2As_4$ are calculated with SOC, as shown in Fig. 3 (a) and (b), respectively. The mainly valley-contrast properties are almost independent of spin, despite of the subtle difference between the results with or without SOC. The bandgap is slightly smaller than the bandgap without SOC, due to spin splitting near the K and K' points. The bandgaps with SOC vary from 2.00 ($MoSi_2N_4$) to 0.41 eV ($MoSi_2As_4$) in K/K' points, covering the visible and infrared range. To our surprise, $MoSi_2As_4$ monolayers have another valley in the second unoccupied band without SOC, namely the third and fourth unoccupied bands with SOC. These additional bands add another degree of freedom in energy in monolayer $MoSi_2As_4$, resulting in a multiple-folded valley. As far as we know, multi-fold valleys have not been reported in any pristine materials.

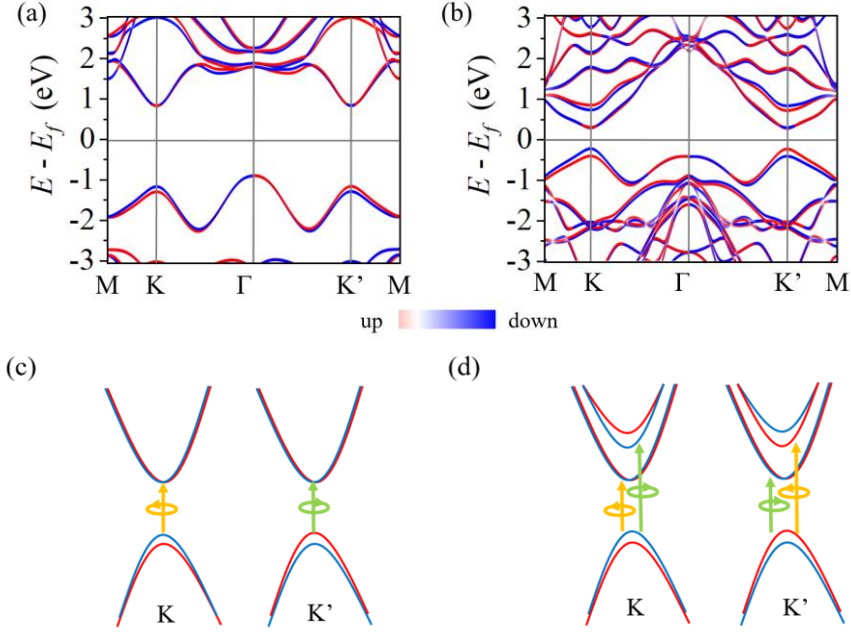

Fig. 3. Fat band structures along the high symmetric lines of (a) monolayer MoSi$_2$N$_4$ and (b) MoSi$_2$As$_4$. Illustrated comparison between (c) traditional valleys and (d) multiple-folded valleys. Circles display the chirality of phonons, and arrow length implies the phonon energy. Red and blue color represents up and down spin states, respectively.

As shown in Fig. 4. (a) and (b), the bands projected on the atomic orbitals demonstrate that the CBM has an angular momentum of zero, and the VBM consists of a state with the angular momentum of $\pm 2$ in monolayer MoSi$_2$As$_4$ at the K/K' point, which is the same to that in monolayer MoSi$_2$N$_4$ at the K/K' point.[34] Besides, the additional valleys in MoSi$_2$As$_4$ comprised of states with an angular momentum of $\mp 2$. The angular momentum is opposite to the ones at VBM.

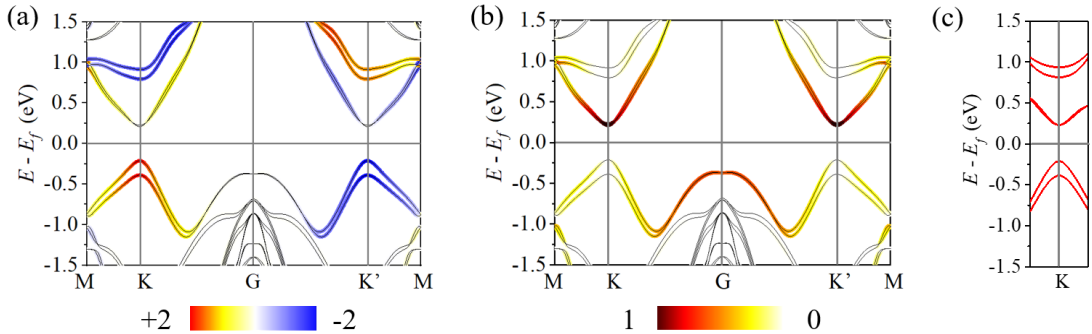

Fig. 4. Fat band structures along the high symmetric lines of monolayer MoSi$_2$As$_4$ with including SOC. (a) Projection on Mo-states with $l = 2$, and $m = \pm 2$. Color represents the angular momentum. (b) Projection on Mo-states with $l = 2$, and $m = 0$. Color represents the weight of projection, respectively. Note that projection on other orbitals is small near the Fermi level at K and K' points. (c) **k · p** band structure near the K point in monolayer MoSi$_2$As$_4$.

By the same method in Eq. (2), we calculated the Berry curvature and circular polarization of the additional valleys of monolayer MoSi$_2$As$_4$, as shown in Fig. 5. Both Berry curvature and circular polarization are also valley contrast. Fig. 5. (a) shows the circular dichroism of photon absorption by exciting an electron from the first occupied band to the second unoccupied bands at the same momentum position. The circular polarization is also valley specified. The selection can even reach 100% for a left-handed or right-handed photon. In the same valley, the circular polarization of a phonon from the first occupied band to the first and second unoccupied bands is also the opposite. Using a lower phonon of near 0.41 eV, we can selectively pump an electron from the CBM to VBM at the K (K') points using a left-handed (right-handed) photon. Using a higher phonon of near 1.00 eV, we can selectively pump an electron from the CBM to addition valleys at the K (K') point by using a right-handed (left-handed) photon. Thus, the additional bands can work as an intrinsic component of a new type of multiple-folded valleys. The optical selection and the comparison between traditional valleys and multiple-folded valleys are shown in Fig. 3 (c) and (d), respectively.

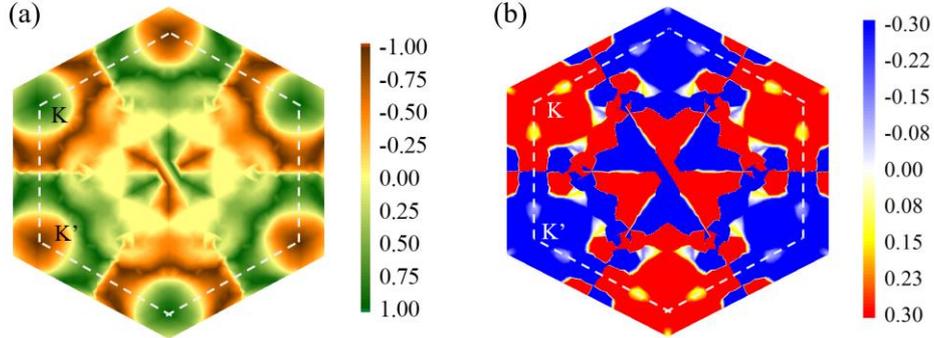

Fig. 5. (a) Circular polarization from the first occupied bands to the second unoccupied band of monolayer MoSi$_2$As$_4$. (b) Berry curvature distribution of the second unoccupied band in the BZ of monolayer MoSi$_2$As$_4$.

Inversion symmetry breaking together with SOC leads to the splitting of about 121 meV at VBM points in monolayer MoSi$_2$N$_4$ and that of 173 meV in monolayer MoSi$_2$As$_4$. Besides, the surface states at both VBM and CBM are spin-polarized in the *z*-direction due to the mirror plane parallel to the material. Spin polarization is also opposite at the K and K' points due to time-reversal symmetry. Hall conductance is proportional to the integration of Berry curvature.

Combing the valley contrast Berry curvature in Fig. 2, the spin Hall can be observed, if suitable holes are injected. Due to approximate zero angular momentum at CBM, the splitting is as small as 16 meV in monolayer MoSi$_2$As$_4$. Similarly, the spin splitting is about 3 meV at the CBM of monolayer MoS$_2$.[38] Due to small spin splitting at CBM in monolayer MoS$_2$, the electron-type spin Hall effect is hard to realize. Fortunately, the additional valleys of monolayer MoSi$_2$As$_4$ have a large spin splitting due to the large angular momentum at K and K' points. Besides, there is also valley-contrast Berry in the second unoccupied band in monolayer MoSi$_2$As$_4$, as shown in Fig. 5. (b). Thus, the electron-type spin Hall effect can be realized in monolayer MoSi$_2$As$_4$.

The carriers residing in the valence band of original valleys are holes, and the carriers in the conduction band of valleys are electrons without spin polarization. The carriers of the additional valleys in monolayer MoSi$_2$As$_4$ are electrons with a polarized spin if weak doped or excited with suitable phonon energy. Thus, these three types of carriers or quasiparticles can be well distinguished by charge and spin in the future experiments of MoSi$_2$As$_4$. These three bands can be an intrinsic dimension of a valley and can be used to store and process multiple information.

To understand the DFT (density functional theory) results, we built a low-power energy model, considering the symmetries of valleys. Although the materials of MoSi$_2$As$_4$ have a spatial group of $D_{3h}$, Hamilton at K and K' has a subgroup of $C_{3h}$. According to the projection calculation using DFT, the periodic part of the wave function can be described by $|d_{z^2}\rangle$ and $|d_{xy} + \tau\xi i d_{x^2-y^2}\rangle$, where $\tau = \pm 1$ is the valley index, and $\xi$ is +1 or -1 for the VBM and the bottom of the second unoccupied band. These three different wave functions belong to three different groups $|A'\rangle$, $|E'_1\rangle$ and $|E'_2\rangle$. These three subgroups have even symmetry under the mirror operator, and only the spin in the z-direction is allowed. Using a similar method in the previous report,[39] we built a $\mathbf{k} \cdot \mathbf{p}$ Hamiltonian in the basis of $|d_{xy} - \tau i d_{x^2-y^2}\rangle \otimes |s_z\rangle$, $|d_{z^2}\rangle \otimes |s_z\rangle$ and $|d_{xy} + \tau i d_{x^2-y^2}\rangle \otimes |s_z\rangle$. We cut up matrix elements to the first order of $\mathbf{k}$, and obtain a three-band $\mathbf{k} \cdot \mathbf{p}$ Hamiltonian contained by $C_{3h}$ symmetry in the following form:

$$H_\tau = \begin{pmatrix} \varepsilon_a + \tau\lambda_a S_z & \gamma_2 k^- & \gamma_1 k^+ \\ \gamma_2 k^+ & \varepsilon_c & \gamma_3 k^- \\ \gamma_1 k^- & \gamma_3 k^+ & \varepsilon_v + \tau\lambda_v S_z \end{pmatrix} \quad (3)$$

where $k^\pm = k_x \pm \tau i k_y$, and $k_x$ or $k_y$ is the momentum measured from the K or K'. $\varepsilon_a$, $\varepsilon_c$ and $\varepsilon_v$ are the energy of K or K' without SOC splitting. $\Delta = \varepsilon_c - \varepsilon_v$ is the bandgap, and $\Delta = \varepsilon_a - \varepsilon_v$ is the energy interval from the top of second unoccupied to VBM. $2\lambda_v$ or $2\lambda_a$ is the spin splitting at VBM or the bottom of the second unoccupied band. $\gamma_1$, $\gamma_2$ and $\gamma_3$ can be obtained by the DFT calculation. Actually, they are the chiral momentum matrix between different bands. In monolayer MoSi$_2$As$_4$, $\gamma_1^2$, $\gamma_2^2$ and $\gamma_3^2$ are 4.6, 198 and 80.5 in the atomic unit, respectively. The $\mathbf{k} \cdot \mathbf{p}$ band structure is in good accordance with DFT calculation, implying our **k·p** Hamiltonian captures the main physics near the K and K' points. For monolayer $MoSi_2N_4$ or $MoS_2$, a two-band Hamiltonian is enough. The **k·p** Hamilton in a similar approximation to $MoSi_2As_4$ is:[40]

$$H_\tau = \begin{pmatrix} \varepsilon_c & \gamma_3 k^- \\ \gamma_3 k^+ & \varepsilon_v + \tau\lambda_c S_z \end{pmatrix} \quad (4)$$

In both Eq. (3) and (4), the spin spitting term and linear dispersion are valley-dependent. The two valleys are connected by time-reversal symmetry, and the valley index changes its sign under the time-reversal operator. The spin splitting is opposite at different valleys, and the lateral spin components are forbidden by the mirror symmetry at the K and K' points. According to $C_3$ symmetry, $C_3|d_{z^2}\rangle = |d_{z^2}\rangle$, and $C_3|d_{xy} + \tau\xi i d_{x^2-y^2}\rangle = e^{\mp i2\pi/3}|d_{xy} + \tau\xi i d_{x^2-y^2}\rangle$. The chiral momentum operator obeys the rule as $C_3 \mathbf{p}_\pm C_3^\dagger = e^{\mp i2\pi/3}\mathbf{p}_\pm$. It is easy to check the momentum matrix of $\mathbf{p}_+$ or $\mathbf{p}_-$ between any two different states. Between any two different states, either of $\mathbf{p}_\pm$ is zero, and the other is nonzero. Thus, circular polarization is perfectly left-handed or right-handed, and whether a left-handed photon or right-handed will be absorbed depends on the phase under $C_3$ rotation.[41] As shown in Fig. 4. (c), the **k·p** band structure near the K point in monolayer $MoSi_2As_4$ is in good accordance to our DFT calculations.

**Conclusion**

In this paper, we predict that the hexagonal monolayer $MoSi_2N_4$ and its derivative $MoSi_2As_4$ have a pair of valleys. The valleys can be controlled by circularly polarized phonons. Due to nonzero and valley contrast Berry curvature, valley Hall and spin Valley can be expected. Beyond traditional two-level valleys, the valleys in monolayer $MoSi_2As_4$ are multiple-folded, implying an additional intrinsic degree of freedom. Each valley and energy band can be selectively controlled. This study presents the multiple-information operator and storage in valleytronic devices.

**Computational Details**

Phonon spectra, band structures, Berry curvature, and optical properties of materials were computed by the Vienna *ab-initio* simulation package (VASP).[42] The projector augmented waves with a cutoff energy of 520 eV was used as the basis set, and the generalized gradient approximation (GGA) combined with the Perdew-Burke-Ernzerhof (PBE) method was used as the exchange and correlation functional.[43-45] To avoid the periodic imagines vertical the layers, a vacuum distance longer than 15 Å remains. The geometry and ions were fully relaxed until the force of each atom decreases to 0.01 eV/Å. The convergence threshold was set as $10^{-6}$ eV for energy. The *k*-point mesh sampled by the Monkhorst-Pack method with a separation of 0.02 Å$^{-1}$ was adopted.

**Acknowledgment**

This work was supported by the Ministry of Science and Technology of China (No. 2016YFB0700600 and 2017YFA206303), the National Natural Science Foundation of China (No. 11674005, 91964101, and 11664026). We also gratefully acknowledged the support from the High-performance Computing Platform of Peking University and the MatCloud + high throughput materials simulation engine.

TOC

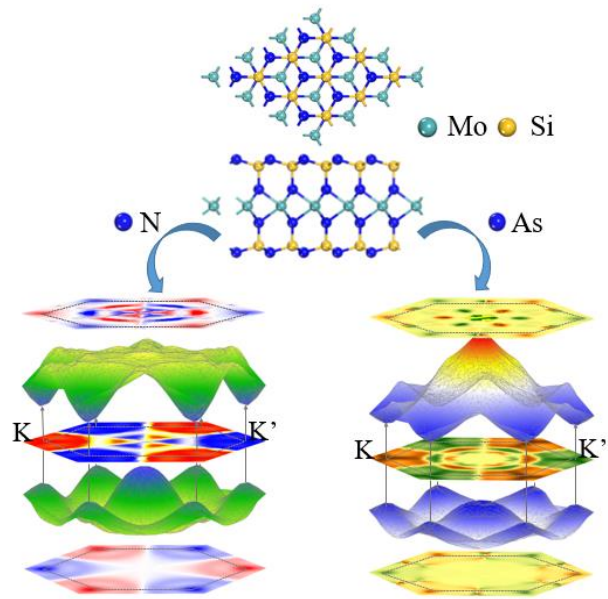

We predict that the valley-contrast properties can be realized in monolayer MoSi$_2$N$_4$ and its derivative MoSi$_2$As$_4$.